\documentclass[prb,twocolumn,amsmath]{revtex4}

\usepackage{graphicx}

\providecommand{\etal}{\emph{et~al.}}
\providecommand{\mysecref}[1]{Sec.~\ref{#1}}
\providecommand{\myeqref}[1]{Eq.~\eqref{#1}}
\providecommand{\myfigref}[1]{Fig.~\ref{#1}}
\providecommand{\mytabref}[1]{Table~\ref{#1}}
\providecommand{\pedex}[1]{\ensuremath{_\text{#1}}}
\providecommand{\apex}[1]{\ensuremath{^\text{#1}}}
\providecommand{\wavef}{\varphi}
\providecommand{\diff}[2][]{\ensuremath{\mathop{\text{d}^{#1}{#2}}}}
\DeclareMathOperator{\bessel}{j}
\providecommand{\deriv}[3][]{\ensuremath{\frac{\diff[#1]{#2}}{\diff{#3}^{#1}}}}
\providecommand{\bra}[1]{\ensuremath{\bigl \langle #1 \bigr \rvert}}
\providecommand{\ket}[1]{\ensuremath{\bigl \lvert #1 \bigr \rangle}}
\providecommand{\braketvec}[2]{\ensuremath{\bigl \langle #1 \bigr \rvert #2 \bigr \rangle}}
\providecommand{\abs}[1]{\ensuremath{\bigl \lvert #1 \bigr \rvert}}
\providecommand{\vect}[1]{\ensuremath{\boldsymbol{#1}}}
\makeatletter
\providecommand{\unit}{\@ifstar\unit@star\unit@nostar}
\providecommand{\unit@star}[1]{\ensuremath{~(\text{#1})}}
\providecommand{\unit@nostar}[1]{\ensuremath{~\text{#1}}}
\makeatother
\providecommand{\fleur}{\texttt{Fleur}}

\usepackage{xcolor}

\begin{document}

\title{Optimization Algorithm for the Generation of ONCV Pseudopotentials}
\author{Martin Schlipf}
\email{martin.schlipf@gmail.com}
\author{Fran\c{c}ois Gygi}
\email{fgygi@ucdavis.edu}
\affiliation{Department of Computer Science, University of California Davis, Davis, CA 95616, USA}

\begin{abstract}
  We present an optimization algorithm to construct
  pseudopotentials and use it to generate a set of
  Optimized Norm-Conserving Vanderbilt (ONCV)
  pseudopotentials for elements up to Z=83 (Bi) 
  (excluding Lanthanides).
  We introduce a quality function that assesses the
  agreement of a pseudopotential calculation with
  all-electron FLAPW results, and the necessary 
  plane-wave energy cutoff. 
  This quality function allows us to use a Nelder-Mead
  optimization algorithm on a training set of materials
  to optimize the input parameters of the pseudopotential
  construction for most of the periodic table. We control
  the accuracy of the resulting pseudopotentials on a test
  set of materials independent of the training set.
  We find that the automatically constructed pseudopotentials
  provide a good agreement with the all-electron results 
  obtained using the FLEUR code with a plane-wave 
  energy cutoff of approximately 60 Ry.
\end{abstract}

\maketitle

\section{Introduction} \label{sec:intro}
Pseudopotentials were introduced over three decades ago as an elegant
simplification of electronic structure computations.\cite{Martin2004}
They allow one to avoid
the calculation of electronic states associated with core electrons, and focus
instead on valence electrons that most often dominate phenomena of interest, in
particular chemical bonding. In the context of Density Functional Theory (DFT),
pseudopotentials have made it possible to solve the Kohn-Sham equations
\cite{hk64,ks65} using a plane-wave basis set, which considerably reduces the
complexity of calculations, and allows for the use of efficient Fast Fourier
Transform (FFT) algorithms. The introduction of norm-conserving
pseudopotentials (NCPP) by Hamann \etal{} in 1979 \cite{hsc79,bhs82} greatly
improved the accuracy of DFT plane wave calculations by imposing a constraint
(norm conservation) in the construction of the potentials, thus improving the
transferability of potentials to different chemical environments. More
elaborate representations of pseudopotentials were later proposed, most notably
ultrasoft pseudopotentials \cite{vand90} (USPP) and the projector augmented
wave \cite{paw} (PAW) method, improving computational efficiency by reducing
the required plane wave energy cutoff. The implementation of these PP is
however rather complex,\cite{hama13}  in particular for advanced calculations
involving hybrid density functionals,\cite{bec2_93}  many-body perturbation
theory,\cite{ag98} or density-functional perturbation theory.\cite{bgt87} Both
USPP and PAW potentials have been used with great success in a large number of
computational studies published over the past two decades. NCPP potentials were
also widely used but suffered from the need to use a large plane wave basis set
for some elements, especially transition metals.

Recently, Hamann suggested\cite{hama13} a method to construct optimized
norm-conserving Vanderbilt (ONCV) potentials following the USPP construction
algorithm without forfeiting the norm-conservation.  The resulting potentials
have an accuracy comparable to the USPP at a moderately increased plane-wave
energy cutoff.

Since the very first pseudopotentials were introduced, there has been an
interest in a database of transferable, reference potentials that could be
applied for many elements in the periodic table.\cite{bhs82,tm91,hgh98} The
need for a systematic database in high-throughput calculations led to a recent
revival of this field: Garrity \etal\cite{gbrv14} proposed a new set of USPP for
the whole periodic table except the noble gases and the rare earths.  Dal
Corso\cite{cors14} constructed a high- and a low-accuracy PAW set for all
elements up to Pu.  Common to these approaches is the fact that the input
parameters of the PP construction are selected by experience based on the
results of the all-electron (AE) calculation of the bare atom.  The quality of
the constructed PP is then tested by an evaluation of different crystal
structures and by comparing to the all-electron FLAPW\cite{wkwf81,wwf82,jf84}
results.  To standardize the testing procedure, Lejaeghere \etal\cite{lsoc14}
suggested to compare the area between a Murnaghan fit\cite{mur44} obtained with
the PP and the AE calculation resulting in a quality factor $\Delta$.
K\"u\c{c}\"ukbenli \etal\cite{kucu14} proposed a crystalline monoatomic solid
test, where the quality factor of the simple cubic (sc), body-centered cubic
(bcc), and face-centered cubic (fcc) structure is evaluated to assess the
quality of a PP.  There are two improvements over these construction principles
that we propose to address in this work.  First, we introduce a quality
function that takes into account the computational efficiency of the PP as well
as its accuracy.  Second, we allow for a systematic feedback of this
quality function onto the input parameters defining the PP.  In this way, we can
introduce an automatic construction algorithm that optimizes the properties of
the PP without bias from the constructor.  We apply this algorithm to
construct ONCV pseudopotentials and compare their performance with recent
USPP\cite{gbrv14} and PAW\cite{cors14} PP libraries. The pseudopotentials
are available in UPF and XML format on our webpage.\cite{webpage}

This paper is organized as follows: In \mysecref{sec:oncv}, we outline the
properties of the ONCV PP and introduce the input parameters that will be
optimized by the algorithm.  In \mysecref{sec:detail}, we introduce the quality
function to assess the performance of the PP, specify the materials we use to
construct and test the PP, outline the setting of the DFT calculation, and
finally present the optimization algorithm that iterates construction and
testing until a good PP is found.  We compare the constructed PP to results
obtained with FLAPW, USPP, and PAW in \mysecref{sec:result} and draw our
conclusions in \mysecref{sec:concl}

\section{ONCV pseudopotentials} \label{sec:oncv}

The optimized norm-conserving Vanderbilt (ONCV) pseudopotentials were recently
proposed by Hamann.\cite{hama13} Here, we briefly sketch their construction,
following Hamann, to highlight the input parameters (bold in text) that
determine the properties of the PP.  The general idea is introduce an
{\bfseries upper limit} wave vector $q\pedex{c}$ and optimize the pseudo wave
functions $\wavef_i(r)$ such that the residual kinetic energy 
\begin{equation}
  E_{ij}(q\pedex c) = \int_{q\pedex{c}}^\infty \diff q q^4 \wavef_i(q) \wavef_j(q)
\end{equation}
above this cutoff is minimized.
Here, $\wavef_i(q)$ is the Fourier transform of the pseudo wave function
\begin{equation}
  \wavef_i(q) = 4\pi \int_0^\infty \diff r r^2 \bessel_l(qr) \wavef_i(r)\medspace,
\end{equation}
$\bessel_l(qr)$ a spherical Bessel function, and $l$ the angular moment of the
pseudo wave function.  On the one hand, the cutoff $q\pedex{c}$ determines
which features of the physical potential can be described by the PP.  On the
other hand, increasing $q\pedex{c}$ makes the PP harder and hence more costly
to evaluate.

For every angular moment, a {\bfseries projector radius} $r\pedex{c}$
determines in which region the pseudoization is done.  The projector radius is
approximately inversely proportional to the cutoff $q\pedex{c}$ so that a
smaller value increases the computational cost along with the accuracy.
Outside of this radius the wave function should follow the true all-electron
wave function $\psi$.  To ensure the continuity at this radius, one imposes $M$
constrains on the continuity of the pseudo wave function

\begin{equation} \label{eq:constr}
  \deriv[n]{\wavef}{r}\biggr|_{r\pedex{c}} = \deriv[n]{\psi}{r}\biggr|_{r\pedex{c}}\medspace,
\end{equation}
for $n = 0, \ldots M-1$. In this work, we use $M=5$ for all constructed PP.

The basis set used in the optimization is constructed from spherical Bessel
functions.  As the basis functions are only used inside the sphere, they are
set to zero outside of the projector radius.  This destroys the orthogonality
of the basis, so that one needs to orthogonalize it again.  A linear
combination of the orthogonalized basis functions yields a new basis where a
single basis function $\wavef_0$ satisfies the constraints in
\myeqref{eq:constr} and for all other basis functions $\xi_n\apex{N}$ the value
and the $M-1$ derivatives at $r\pedex{c}$ are zero.  As a consequence, the sum
of $\wavef_0$ and any linear combination of the $\xi_n\apex{N}$ will satisfy
the constraints in \myeqref{eq:constr}.  It is advantageous to select those
linear combinations of $\xi_n\apex{N}$ that have a maximal impact on the
residual energy by evaluating the eigenvalues $e_n$ and eigenvectors
$\xi_n\apex{R}$
\begin{equation}
  \wavef_i = \wavef_0 + \sum_{n=1}^{N-M} x_n \xi_n\apex{R}\medspace.
\end{equation}
In this work, we construct the PP with $N = 8$ basis functions.  Notice that
the optimization of the pseudo wave function is performed under the constraint
that the norm of the all-electron wave function is conserved.

From the obtained pseudo wave functions, one can construct projectors $\chi_i$
\begin{equation}
  \chi_i(r) = (\varepsilon_i - T - V\pedex{loc})\phi_i(r)\medspace,
\end{equation}
where $T$ is the kinetic energy operator.  $V\pedex{loc}$ is the local
potential that follows the all-electron potential outside of $r\pedex{c}$ and
is extended smoothly to the origin by a polynomial.  For occupied states
$\varepsilon_i$ is the eigenvalue of the all-electron calculation.  For
unoccupied states, one needs to specify this {\bfseries energy shift} before
the construction of the PP.  Following Ref.~\onlinecite{hama13}, we 
construct two projectors per angular momentum $l \le l\pedex{max}$ and only the
local potential for all $l > l\pedex{max}$ above.  The projectors define the
following nonlocal potential
\begin{equation}
  V\pedex{NL} = \sum_{ij} \ket{\chi_i}B_{ij}^{-1}\bra{\chi_j}
\end{equation}
where 
\begin{equation}
  B_{ij} = \braketvec{\wavef_i}{\chi_j}\medspace,
\end{equation}
which is a Hermitian matrix when normconserving pseudo wave functions are
constructed.\cite{vand90} One can simplify this potential by a unitary
transformation to the eigenspace of the $B$ matrix.

\section{Computational Details} \label{sec:detail}

\subsection{Quality function}

\begin{figure}
  \includegraphics{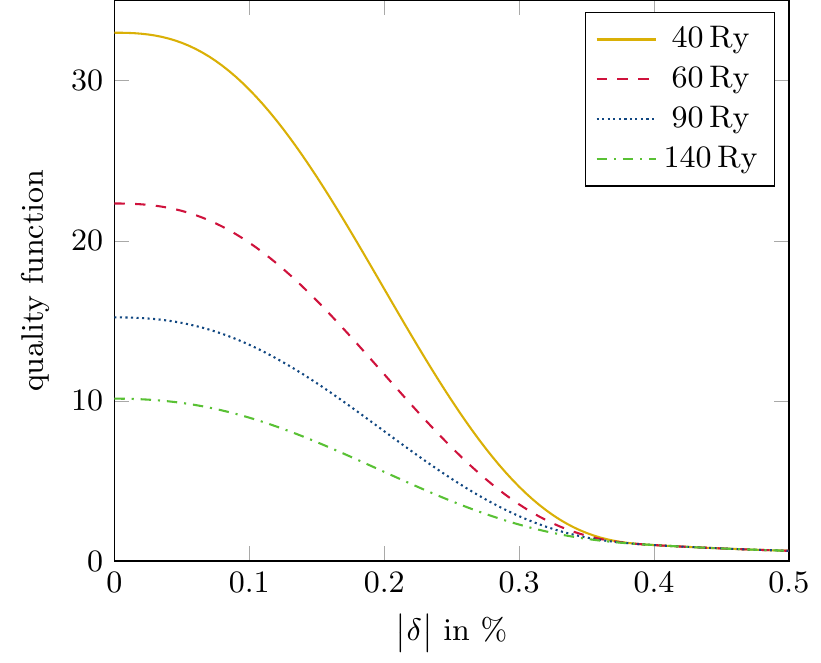}
  \centering
  \caption{(Color online) Quality function for various energy-cutoffs $E\pedex{cut}$. For small $\delta$, it
    is proportional to $1/E\pedex{cut}$; for large $\delta$ proportional to $1/\delta^2$ and independent of
    $E\pedex{cut}$.}
  \label{fig:quality}
\end{figure}
In order to employ numerical optimization algorithms in the construction of
PPs, we need a function that maps the multidimensional input parameter space
onto a single number, the \emph{quality} of the PP.  A good PP is characterized
by a small relative deviation 
\begin{equation}
  \delta\apex{PP}_{a\pedex{lat}} = a\apex{PP}\pedex{lat} / a\apex{AE}\pedex{lat} - 1
\end{equation}

between the lattice constant obtained in the plane-wave PP calculation
$a\pedex{lat}\apex{PP}$ and in the AE calculation $a\pedex{lat}\apex{AE}$,
respectively.  A second criterion is the plane-wave energy cutoff
$E\pedex{cut}$ necessary to converge the PP calculation.  These two criteria
compete with each other because the pseudoization of the potential reduces the
necessary energy cutoff at the cost of a lower accuracy near the nucleus.
Hence, we need to specify a target accuracy $\delta_0$ which we want to achieve
for our PP, i.e., for all materials $\abs{\delta\apex{PP}_{a\pedex{lat}}} \le
\delta_0$.  We select $\delta_0 = 0.2\%$ motivated by the fact that the choice
of different codes or input parameters in the all-electron calculation may
already lead to a relative error of approximately $0.1\%$.  To discriminate
between PPs within the target accuracy, we include a term $\propto
1/E\pedex{cut}$ in the quality function, favoring smoother PP over hard ones.
For PPs that are significantly outside $\abs{\delta\apex{PP}_{a\pedex{lat}}} >
2\delta_0$ our target accuracy, we only focus on optimizing the relative
deviation by an $1/(\delta\apex{PP}_{a\pedex{lat}})^2$ term.  We choose a
smooth continuation between the two regions, resulting in the function depicted
in \myfigref{fig:quality}.  The quality function has the following form
\begin{equation}
  q(\delta, E\pedex{cut}) = \begin{cases}
    A + C \delta^2 + D \delta^3 + E \delta^4 + F \delta^5 & 
  \delta < 2\delta_0 \\
    \left(2\delta_0  / \delta\right)^2 & \delta \ge 2\delta_0
  \end{cases}
\end{equation}
with
\begin{align}
  A &= 1 + \frac{1280}{E\pedex{cut}} & 
  y_0 &= 1 + \frac{680}{E\pedex{cut}} \nonumber \\ 
  C &= \frac{32 y_0 - 16 A -29}{4\delta_0^3} & 
  D &= \frac{19 A - 48 y_0  + 54}{4 \delta_0^2} \nonumber \\
  E &= \frac{96 y_0 - 33 A - 122}{16 \delta_0^4} & 
  F&= \frac{5A -16y_0 +22}{16\delta_0^5}\medspace. \nonumber
\end{align}
The function can be multiplied by an arbitrary scaling constant, which we set
such that the value of the quality function is 1 at
$\abs{\delta\apex{PP}_{a\pedex{lat}}} = 2\delta_0$.

\subsection{Sets of materials}

As the constructed pseudopotentials depend on the set of materials used in the
optimization algorithm, it is important that the set contain physically
relevant environments of the atom.  Furthermore, we select highly symmetric
structures with at most two atoms per unit cell to reduce the computation time.
As representatives of a metallic environment, we select the simple cubic (sc),
the body-centered cubic (bcc), the face-centered cubic (fcc), and the
diamond-cubic (dc) structure.  Ionic environments are provided in a rock-salt
or zinc-blende structure, where we combine elements such that they assume their
most common oxidation state.  This leads to a combination of elements from the
lithium group with the fluorine group, the beryllium group with the oxygen
group, and so on.  We always use the three smallest elements of the respective
groups to guarantee a variation in size of the resulting compounds.  For the
transition metals, several oxidation states are often possible.  Hence, we
combine them with carbon, nitrogen, and oxygen to test these different
valencies.  As the noble gases do not form compounds, we test them only in the
sc, bcc, fcc, and dc structure. 

Finally, we need to separate these materials into two sets.  The \emph{training
set} consists of the bcc, and the fcc structure as well as all rock-salt
compounds.  It is used in the optimization algorithm to construct the PPs.  As
the PPs are specifically optimized to reproduce the structural properties of
the training set, we can only judge if the PPs are highly accurate by
calculating an independent \emph{test set}.  The test set contains the sc and
the dc structure as well as all zinc-blende compounds.  In total, the training
and test sets consist of 602 materials, where every noble-gas atom is part of
four materials, and every other element is part of at least ten materials.

\subsection{Computational setup}

All pseudopotentials are constructed using the Perdew-Burke-Ernzerhof (PBE)
generalized gradient density functional.\cite{pbe96} We use an $8 \times 8
\times 8$ Monkhorst-Pack $\vect k$-point mesh in the AE as well as in the PP
calculation.  While this may not be sufficient to completely converge the
lattice constant with respect to the numbers of $\vect k$-points, the errors in
the PP and the AE calculation are expected to be the same, so that we can still
compare the results.  To ensure that metallic systems converge, we use a
Fermi-type smearing with a temperature of $315.8\unit{K}$ corresponding to an
energy of $0.001\unit{htr}$.
 
For the AE calculation, we use the FLAPW method as implemented in the
\fleur{} code.\cite{fleur} We converge the plane-wave cutoff and add unoccupied
local orbitals to provide sufficient variational freedom inside the muffin-tin
spheres.  The precise numerical values necessary to converge the calculation
are different for every material; all input files can be
obtained from our web page.\cite{webpage} We obtain the lattice constant by a
Murnaghan fit\cite{mur44} through 11 data points surrounding the minimum of the
total energy.

\begin{figure}
  \includegraphics{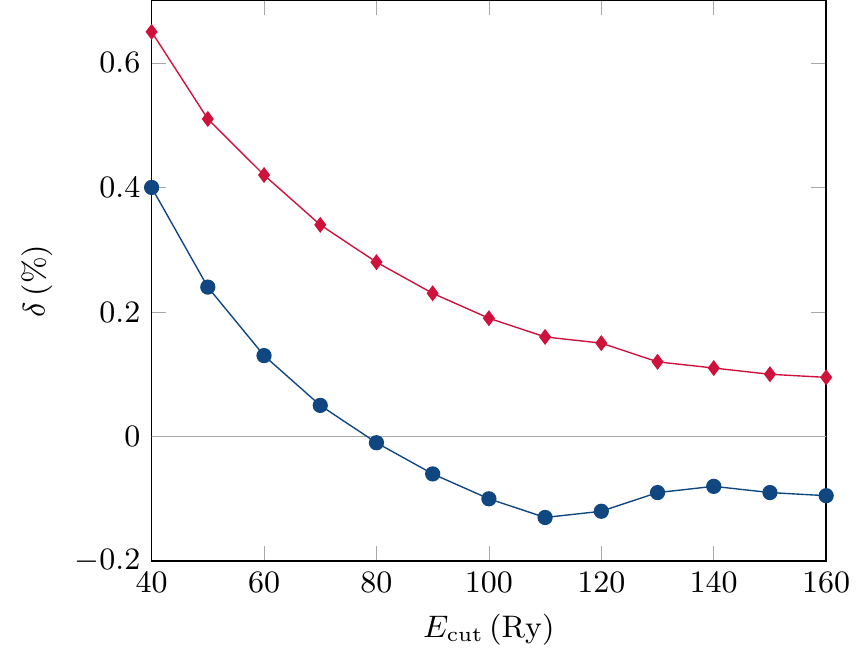}
  \centering
  \caption{(Color online) Relative deviation $\delta$ of a PP 
  w.r.t. the AE calculation.  The blue circles indicate the deviation 
  obtained at a certain energy cutoff $E\pedex{cut}$.
  The red diamonds show the corrected deviation that is monotonically
  decreasing with increasing cutoff (see text).}
  \label{fig:err_corr}
\end{figure}
The automatic construction of pseudopotentials requires every material to be
calculated several hundred times.  Hence, we approximate the Murnaghan equation
of state by a parabola that we fit through data points at the AE lattice
constant and a 1\% enlarged or reduced value.  We test the constructed PP with
the {\sc Quantum ESPRESSO}\cite{QE2009} plane-wave DFT code.  Our test consists
of a calculation with a large energy cutoff of $E\pedex{cut}\apex{max} =
160\unit{Ry}$ that we consider to be the converged solution.  Then, we decrease
the cutoff in steps of $\Delta E = 10\unit{Ry}$ to the minimum of
$40\unit{Ry}$.  Notice that as illustrated by \myfigref{fig:err_corr}, the
actual deviation compared to the AE calculation may decrease even though we
reduced the accuracy of the calculation.  To correct for this, we adjust the
deviation such that it is monotonically decreasing using the following 
correction

\begin{equation}
  \delta\apex{PP}\pedex{corr}(E\pedex{cut}^i) = 
  |\delta\apex{PP}(E\pedex{cut}\apex{max})|
  + \sum_{k=1}^i | 
  \delta\apex{PP}(E\pedex{cut}^k) -
  \delta\apex{PP}(E\pedex{cut}^{k-1})|
\end{equation}
where $E\pedex{cut}^i = E\pedex{cut}\apex{max} - 10 i $. 
This ensures that the deviation at a given cutoff energy is an upper
bound to the deviation at any larger cutoff.

\subsection{Optimizing pseudopotentials}

We start from a reasonable starting guess for the $N$ input parameters.  We
used the example input files provided with the ONCVPSP package,\cite{hama13}
where available, or generated our own PP otherwise.  By randomly altering all
input parameters in the starting-guess PP by a small amount, we can construct
new PP.  We assess these PP by evaluating the quality function on the training
set of materials with the geometric average of all involved materials.  In the
case of the rock-salt compounds, we test always only one of the PP and for the
other element we use a PP from the GBRV database.\cite{gbrv14} After $(N+1)$ PP
have been constructed, we employ a Nelder-Mead algorithm\cite{nm65} to optimize
the PP further.  The PP parameters form a simplex in an $N$ dimensional space.
By replacing the worst corner by a better PP the simplex contracts towards the
optimal PP.  The advantages of the Nelder-Mead algorithm are that we do not
need to know the derivatives with respect to the input parameters and that it
can find PP parameters that lie outside of the starting simplex.

After 80 to 200 iterations of the Nelder-Mead algorithm, all PP have converged.
Then, we restart the algorithm using these first generation PP as starting
guess.  Now, we employ the first generation PP in the compounds so that our
resulting PP become independent of the GBRV database.  Once the second
generation is converged as well (another 100 iterations), the properties of the
training set are well reproduced for almost all materials.

\subsection{Refining the training set}

For a few materials, the second generation PP do not reproduce the AE results
on the test set of materials.  Our proposed optimization algorithm provides an
easy solution to overcome these cases by adding additional materials to the
training set.  In particular, for the early transition metals (Sc to Mn) it is
necessary to include the sc structure in the training set.  Furthermore, we
include the dimer of hydrogen and nitrogen into the test set, because the
second generation PPs for these two elements do not describe the bond length of
the dimer accurately.

We emphasize that our optimization algorithm could account for other material
properties.  As long as one is able to define a quality function, which maps
the result of a PP potential calculation onto a number, it is possible to
optimize the input parameters of the PP generation by standard numerical
optimization techniques.

\section{Results} \label{sec:result}

We compare the performance of the ONCV PP constructed in this work
(SG15)\cite{webpage} with the USPP in the GBRV database\cite{gbrv14} and the
high-accuracy PAW in the PSLIB.\cite{cors14} For the latter, we generate the
potentials of PSLIB version 1.0.0 with {\sc Quantum ESPRESSO} version 5.1.1.
In the first subsection, we focus on the lattice constants and bulk moduli of
the materials in the training set.  In the second subsection, we investigate
the materials in the test set.  In the third subsection, we look into materials
that are not represented in the test set to check the accuracy of the
pseudopotentials. In the first two subsections, we focus only on the trends
across all materials in the training and test set, respectively. 

\subsection{Training set}

\begin{table}
  \caption{Comparison of the performance of the USPP in the 
  GBRV database\cite{gbrv14} and the high-accuracy PAW in PSLIB\cite{cors14}
  with the ONCV PP in the SG15 database (this work) for materials 
  in a bcc structure. We analyze the relative deviation of the lattice 
  constant $\delta_{a\pedex{lat}}$ and the bulk modulus $\delta_{B_0}$ 
  between a PP and the AE calculation. The average reveals if the PP 
  has a systematic bias and the root-mean-square (rms) average tests 
  the size of the error. We also show the proportion of materials that are not
  accurately described at various energy cutoffs.}
  \label{tab:result-bcc}
  \begin{ruledtabular}
    \begin{tabular}{c c c c}
      & GBRV & PSLIB & SG15 \\
      \hline
      average $\delta_{a\pedex{lat}}$ (\%)     &   0.03 &   0.03 &   0.04 \\
      rms average $\delta_{a\pedex{lat}}$ (\%) &   0.12 &   0.11 &   0.08 \\
      \% of materials with $|\delta_{a\pedex{lat}}| >  0.2\%$\footnotemark[1] &  10.94 &  40.00 &  23.19 \\
      \% of materials with $|\delta_{a\pedex{lat}}| >  0.2\%$\footnotemark[2] &   9.38 &  15.56 &   8.70 \\
      \% of materials with $|\delta_{a\pedex{lat}}| >  0.2\%$\footnotemark[3] &   9.38 &   4.44 &   2.90 \\
      \hline
      average $\delta_{B_0}$ (\%)     &   0.36 &  -0.32 &   0.52 \\
      rms average $\delta_{B_0}$ (\%) &   3.31 &   2.53 &   3.19 \\
      \% of materials with $|\delta_{B_0}| >  5.0\%$\footnotemark[1] &  25.00 &  62.22 &  53.62 \\
      \% of materials with $|\delta_{B_0}| >  5.0\%$\footnotemark[2] &  14.06 &  26.67 &  18.84 \\
      \% of materials with $|\delta_{B_0}| >  5.0\%$\footnotemark[3] &   9.38 &   8.89 &   7.25 \\
      \hline
      total number of materials &     64 &     45 &     69 \\
    \end{tabular}
    \footnotetext[1]{With an energy cutoff of $40 \text{Ry}$.}
    \footnotetext[2]{With an energy cutoff of $60 \text{Ry}$.}
    \footnotetext[3]{With an energy cutoff of $160 \text{Ry}$.}
  \end{ruledtabular}
\end{table}

\begin{figure*}
\includegraphics[scale=0.78,angle=90]{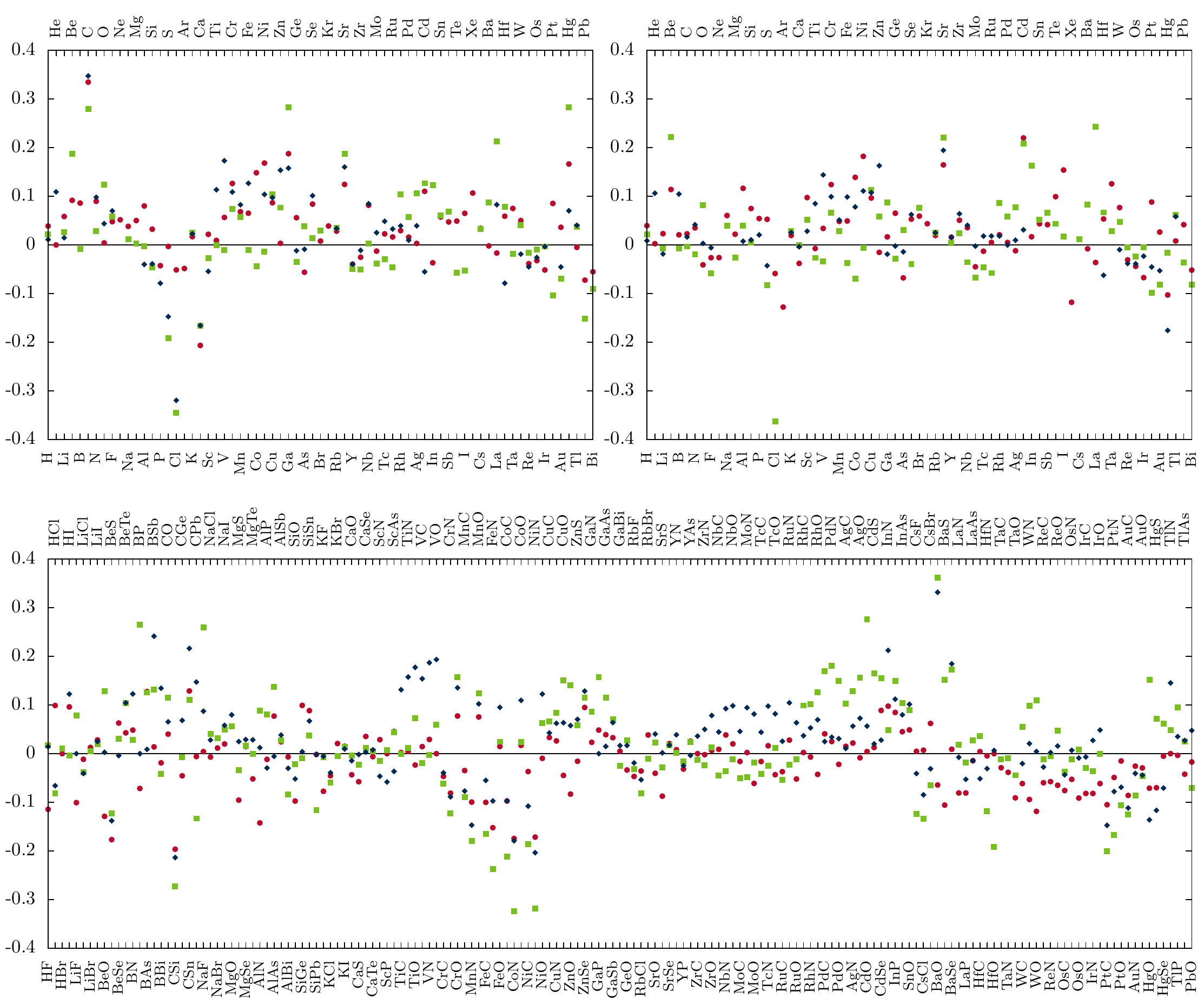}
\caption{(Color online) Relative change $\delta (\%)$ of the lattice 
constant in the training set for the SG15 (red circle), the 
GBRV (green square), and the PSLIB (blue diamond) results as compared 
to the FLAPW ones for the bcc (top left), fcc (top right) and 
rock-salt compounds (bottom).}
\label{fig:training}
\end{figure*}

In \mytabref{tab:result-bcc}, we present the results obtained for the materials
in a bcc structure.  We see that the USPP potentials require the smallest
energy cutoff and have the best performance at $40\unit{Ry}$.  On the other
hand increasing the energy cutoff beyond $40\unit{Ry}$ hardly improves the
results.  For the PAW and the ONCV PP, a large number of materials are not
converged at $40\unit{Ry}$, but increasing the energy cutoff improves the
accuracy, so that they are able to improve on the USPP results.  For the
converged calculation, the root-mean-square (rms) error is around $0.1\%$ for
all PP and smallest for the ONCV PP.  We see a similar trend for the bulk
moduli though the converged results require a larger energy cutoff on average.
The average error for the converged bulk moduli is roughly $3\%$ and the USPP
potentials converge with a lower energy cutoff than the PAW and the ONCV PP,
which have a similar convergence behavior.  In \myfigref{fig:training}, we see
that only for two materials (carbon and calcium) does the converged lattice
constant deviate by more than $0.2\%$ with the ONCV PP. For both of these
materials the USPP and the PAW approach show large deviations as well.

\begin{table}
  \caption{Same as \mytabref{tab:result-bcc} for fcc structures.}
  \label{tab:result-fcc}
  \begin{ruledtabular}
    \begin{tabular}{c c c c}
      & GBRV & PSLIB & SG15 \\
      \hline
      average $\delta_{a\pedex{lat}}$ (\%)     &   0.03 &   0.03 &   0.03 \\
      rms average $\delta_{a\pedex{lat}}$ (\%) &   0.11 &   0.07 &   0.07 \\
      \% of materials with $|\delta_{a\pedex{lat}}| >  0.2\%$\footnotemark[1] &   9.38 &  27.08 &  24.64 \\
      \% of materials with $|\delta_{a\pedex{lat}}| >  0.2\%$\footnotemark[2] &   9.38 &   6.25 &   5.80 \\
      \% of materials with $|\delta_{a\pedex{lat}}| >  0.2\%$\footnotemark[3] &   9.38 &   0.00 &   1.45 \\
      \hline
      average $\delta_{B_0}$ (\%)     &   0.23 &   0.00 &   0.31 \\
      rms average $\delta_{B_0}$ (\%) &   2.28 &   1.83 &   2.00 \\
      \% of materials with $|\delta_{B_0}| >  5.0\%$\footnotemark[1] &  12.50 &  68.75 &  43.48 \\
      \% of materials with $|\delta_{B_0}| >  5.0\%$\footnotemark[2] &   7.81 &  16.67 &  17.39 \\
      \% of materials with $|\delta_{B_0}| >  5.0\%$\footnotemark[3] &   3.12 &   4.17 &   5.80 \\
      \hline
      total number of materials &     64 &     48 &     69 \\
    \end{tabular}
    \footnotetext[1]{With an energy cutoff of $40 \text{Ry}$.}
    \footnotetext[2]{With an energy cutoff of $60 \text{Ry}$.}
    \footnotetext[3]{With an energy cutoff of $160 \text{Ry}$.}
  \end{ruledtabular}
\end{table}

The fcc structures presented in \mytabref{tab:result-fcc} follow the same trend
as the bcc structures.  The USPP potentials require the smallest energy cutoff
but can not be improved further by increasing the energy cutoff.  The PAW and
the ONCV PP require a energy cutoff of $60\unit{Ry}$ to converge most
materials, but have fewer inaccurate elements when increasing the energy
cutoff.  Overall the ONCV PP and the PAW are a bit better than the USPP, but
all PP are close to the AE results.  In \myfigref{fig:training}, we see that
only a single material (cadmium) is outside the $0.2\%$ boundary, when using
the converged calculation and the ONCV PP.  The USPP result shows a deviation
of similar size for this material, whereas the PAW lattice constant is close to
the FLAPW result.
 
\begin{table}
  \caption{Same as \mytabref{tab:result-bcc} for rocksalt structures.}
  \label{tab:result-rocksalt}
  \begin{ruledtabular}
    \begin{tabular}{c c c c}
      & GBRV & PSLIB & SG15 \\
      \hline
      average $\delta_{a\pedex{lat}}$ (\%)     &   0.01 &   0.02 &  -0.02 \\
      rms average $\delta_{a\pedex{lat}}$ (\%) &   0.11 &   0.09 &   0.06 \\
      \% of materials with $|\delta_{a\pedex{lat}}| >  0.2\%$\footnotemark[1] &   6.13 &  35.95 &  30.67 \\
      \% of materials with $|\delta_{a\pedex{lat}}| >  0.2\%$\footnotemark[2] &   6.13 &   6.54 &   1.23 \\
      \% of materials with $|\delta_{a\pedex{lat}}| >  0.2\%$\footnotemark[3] &   6.13 &   3.92 &   0.00 \\
      \hline
      average $\delta_{B_0}$ (\%)     &  -0.02 &  -0.03 &  -0.48 \\
      rms average $\delta_{B_0}$ (\%) &   1.67 &   1.29 &   1.34 \\
      \% of materials with $|\delta_{B_0}| >  5.0\%$\footnotemark[1] &   1.84 &  53.59 &  55.83 \\
      \% of materials with $|\delta_{B_0}| >  5.0\%$\footnotemark[2] &   1.84 &   5.23 &  10.43 \\
      \% of materials with $|\delta_{B_0}| >  5.0\%$\footnotemark[3] &   1.84 &   0.00 &   0.61 \\
      \hline
      total number of materials &    163 &    153 &    163 \\
    \end{tabular}
    \footnotetext[1]{With an energy cutoff of $40 \text{Ry}$.}
    \footnotetext[2]{With an energy cutoff of $60 \text{Ry}$.}
    \footnotetext[3]{With an energy cutoff of $160 \text{Ry}$.}
  \end{ruledtabular}
\end{table}

When combining two materials to form rock-salt compounds, we obtain the results
depicted in \mytabref{tab:result-rocksalt}.  In comparison to the metallic (bcc
and fcc) system, the accuracy for the ionic compounds is a bit higher in
particular for the bulk modulus.  With a large energy cutoff the ONCV PPs
essentially reproduce the AE results and the accuracy at $60\unit{Ry}$ for the
lattice constant is very good.  For the bulk modulus, about $10\%$ of the
materials require a larger energy cutoff.  The USPPs have a slightly larger
mismatch for the lattice constants, but converge both lattice constants and
bulk moduli with $40\unit{Ry}$.  The PAW potentials provide a similar
convergence behavior as the ONCV potentials; they deviate a bit more for the
lattice constants, but provide slightly better bulk moduli.

\begin{figure}
  \includegraphics{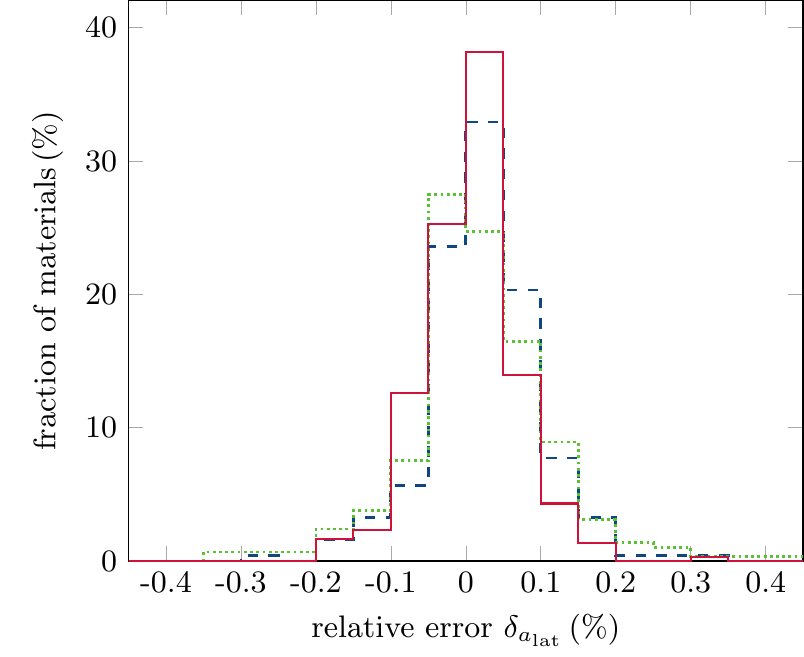}
  \centering
  \caption{(Color online) Histogram of the relative error of the lattice
constant compared to the all-electron result. We show the results for all
materials in the training set for the SG15 (red solid line), the GBRV (green
dotted line), and the PSLIB (blue dashed line) calculations.}
  \label{fig:histo-training}
\end{figure}
In \myfigref{fig:histo-training}, we show a histogram of the relative error of
the lattice constant for all the examined PP (with the converged cutoff of
$160\unit{Ry}$).  The histogram confirms the conclusions we drew from
\mytabref{tab:result-bcc} to \ref{tab:result-rocksalt}: All PPs show a very good
agreement with the all-electron results and the USPPs have a slightly lower
accuracy.  The tails with large errors are very flat indicating that there are
only a few outliers.

\subsection{Test set}
\begin{table}
  \caption{Same as \mytabref{tab:result-bcc} for sc structures.}
  \label{tab:result-sc}
  \begin{ruledtabular}
    \begin{tabular}{c c c c}
      & GBRV & PSLIB & SG15 \\
      \hline
      average $\delta_{a\pedex{lat}}$ (\%)     &   0.02 &   0.03 &   0.02 \\
      rms average $\delta_{a\pedex{lat}}$ (\%) &   0.12 &   0.09 &   0.09 \\
      \% of materials with $|\delta_{a\pedex{lat}}| >  0.2\%$\footnotemark[1] &   6.25 &  46.30 &  27.54 \\
      \% of materials with $|\delta_{a\pedex{lat}}| >  0.2\%$\footnotemark[2] &   6.25 &  16.67 &   5.80 \\
      \% of materials with $|\delta_{a\pedex{lat}}| >  0.2\%$\footnotemark[3] &   6.25 &   3.70 &   2.90 \\
      \hline
      average $\delta_{B_0}$ (\%)     &   0.32 &   0.31 &  -0.01 \\
      rms average $\delta_{B_0}$ (\%) &   3.79 &   3.96 &   4.47 \\
      \% of materials with $|\delta_{B_0}| >  5.0\%$\footnotemark[1] &  40.62 &  74.07 &  62.32 \\
      \% of materials with $|\delta_{B_0}| >  5.0\%$\footnotemark[2] &  20.31 &  27.78 &  21.74 \\
      \% of materials with $|\delta_{B_0}| >  5.0\%$\footnotemark[3] &  12.50 &  12.96 &  11.59 \\
      \hline
      total number of materials &     64 &     54 &     69 \\
    \end{tabular}
    \footnotetext[1]{With an energy cutoff of $40 \text{Ry}$.}
    \footnotetext[2]{With an energy cutoff of $60 \text{Ry}$.}
    \footnotetext[3]{With an energy cutoff of $160 \text{Ry}$.}
  \end{ruledtabular}
\end{table}

\begin{figure*}
  \includegraphics[scale=0.78,angle=90]{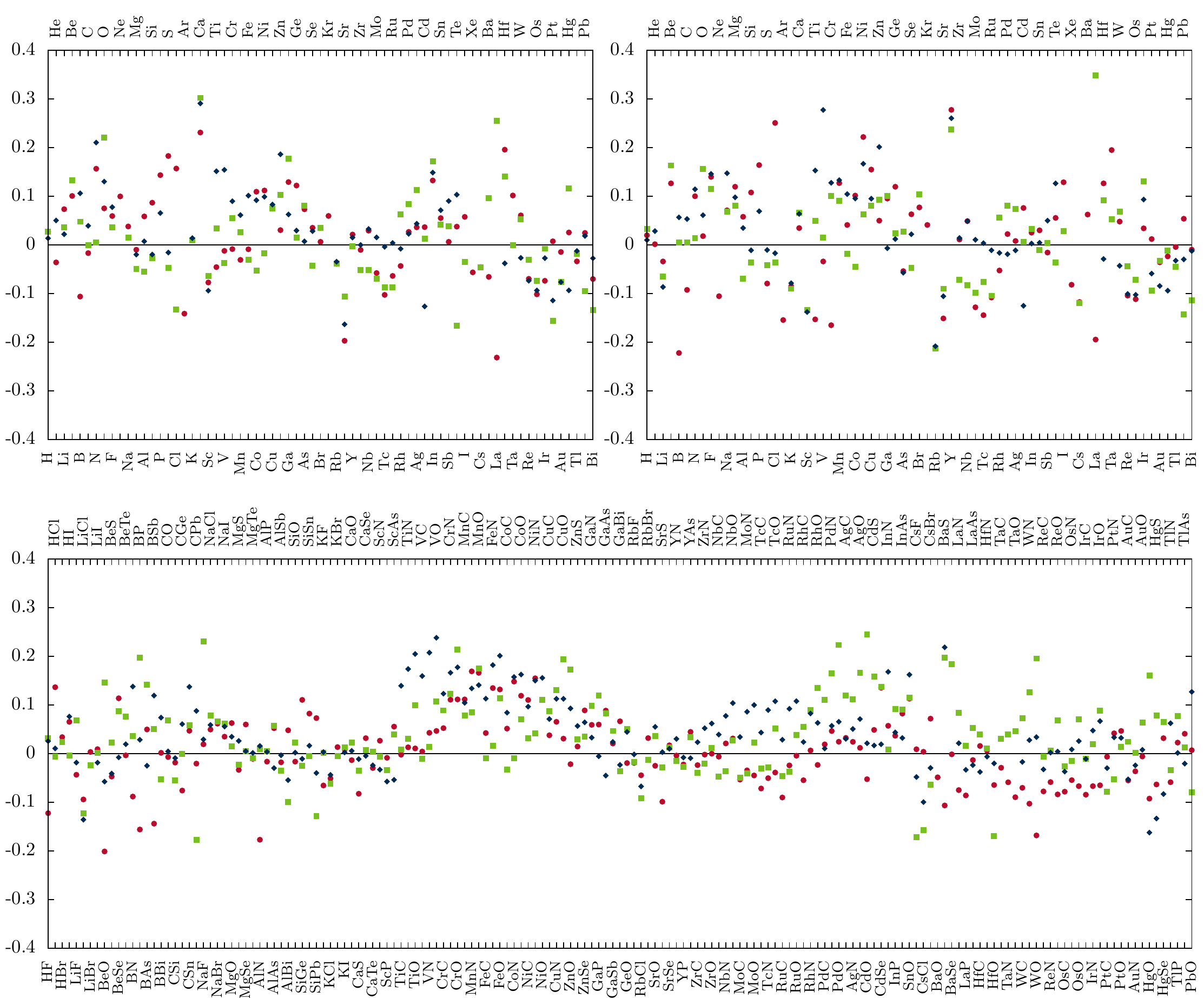}

  \caption{(Color online) Relative change $\delta (\%)$ of the lattice constant
in the test set for the SG15 (red circle), the GBRV (green square), and the
PSLIB (blue diamond) results as compared to the FLAPW ones for the sc (top
left), diamond (top right) and zincblende compounds (bottom).}

\label{fig:test}
\end{figure*}

In the sc structure (see \mytabref{tab:result-sc}), the performance of the ONCV
potentials is comparable to the training set for the lattice constants and
slightly worse for the bulk moduli.  We observe the same trend also for the
USPP and the PAW calculations.  With an overall deviation of about $0.1\%$ for
the lattice constant and $4\%$ for the bulk moduli, all PPs show a good
agreement with the AE reference data.  The convergence with respect to the
energy cutoff is best in the GBRV database, which does not change significantly
for the lattice constants above $40\unit{Ry}$.  Most of the ONCV lattice
constants converge at $60\unit{Ry}$ whereas PAW occasionally needs a larger
cutoff.  For the bulk moduli, all PPs show a similar convergence behavior.
However, we observe that as compared to the other structures a larger fraction
of $>10\%$ is not accurate even with an energy cutoff of $160\unit{Ry}$.  In
\myfigref{fig:test}, we see that the ONCV PPs reproduce the lattice constant
within the $0.2\%$ boundary for all materials except calcium and lanthanum.
While the ONCV PP gives similar results to the other PP for calcium, we find
that the lattice constant in lanthanum is underestimated by the ONCV PP and
overestimated by the USPP.  For this material, the PAW calculation did not
converge.

\begin{table}
  \caption{Same as \mytabref{tab:result-bcc} for diamond structures.}
  \label{tab:result-diamond}
  \begin{ruledtabular}
    \begin{tabular}{c c c c}
      & GBRV & PSLIB & SG15 \\
      \hline
      average $\delta_{a\pedex{lat}}$ (\%)     &   0.03 &   0.02 &   0.01 \\
      rms average $\delta_{a\pedex{lat}}$ (\%) &   0.16 &   0.10 &   0.12 \\
      \% of materials with $|\delta_{a\pedex{lat}}| >  0.2\%$\footnotemark[1] &   7.81 &  49.12 &  34.78 \\
      \% of materials with $|\delta_{a\pedex{lat}}| >  0.2\%$\footnotemark[2] &   7.81 &  22.81 &  11.59 \\
      \% of materials with $|\delta_{a\pedex{lat}}| >  0.2\%$\footnotemark[3] &   7.81 &   7.02 &   8.70 \\
      \hline
      average $\delta_{B_0}$ (\%)     &   0.45 &   1.30 &  -0.24 \\
      rms average $\delta_{B_0}$ (\%) &   4.49 &   6.54 &   3.06 \\
      \% of materials with $|\delta_{B_0}| >  5.0\%$\footnotemark[1] &  31.25 &  71.93 &  53.62 \\
      \% of materials with $|\delta_{B_0}| >  5.0\%$\footnotemark[2] &  18.75 &  31.58 &  14.49 \\
      \% of materials with $|\delta_{B_0}| >  5.0\%$\footnotemark[3] &   9.38 &   7.02 &   7.25 \\
      \hline
      total number of materials &     64 &     57 &     69 \\
    \end{tabular}
    \footnotetext[1]{With an energy cutoff of $40 \text{Ry}$.}
    \footnotetext[2]{With an energy cutoff of $60 \text{Ry}$.}
    \footnotetext[3]{With an energy cutoff of $160 \text{Ry}$.}
  \end{ruledtabular}
\end{table}

In \mytabref{tab:result-diamond}, we present our results for the materials in
the diamond structure.  These are the structures which exhibit overall the
largest deviation from the all-electron result.  The lattice constants of the
USPPs are converged well with the energy cutoff of $40\unit{Ry}$, whereas the
PAW and the ONCV PP frequently require a cutoff of $60\unit{Ry}$.  For the bulk
moduli, we find that the ONCV PP provide the best agreement with the AE
results.  The quality of the USPP is similar, but the PAW potentials show an
average error larger than the desired $5\%$ tolerance.  However the fraction
of materials that are well described with the PP calculation is similar for all
methods.  This indicates that a few specific materials show a particular large
deviation, whereas the rest is accurately described.  For the ONCV PPs the
lattice constants of boron, chlorine, scandium, nickel, rubidium, and yttrium
deviate by more than $0.2\%$ from the FLAPW results.  In \myfigref{fig:test},
we observe that the deviations between the different pseudoizations are larger
than for the other structures.  A possible explanation is that the diamond
structure is a extreme case for many materials, because of its low space
filling.

\begin{table}
  \caption{Same as \mytabref{tab:result-bcc} for zincblende structures.}
  \label{tab:result-zincblende}
  \begin{ruledtabular}
    \begin{tabular}{c c c c}
      & GBRV & PSLIB & SG15 \\
      \hline
      average $\delta_{a\pedex{lat}}$ (\%)     &   0.04 &   0.04 &   0.00 \\
      rms average $\delta_{a\pedex{lat}}$ (\%) &   0.10 &   0.09 &   0.07 \\
      \% of materials with $|\delta_{a\pedex{lat}}| >  0.2\%$\footnotemark[1] &   5.52 &  37.50 &  33.74 \\
      \% of materials with $|\delta_{a\pedex{lat}}| >  0.2\%$\footnotemark[2] &   4.91 &   6.58 &   2.45 \\
      \% of materials with $|\delta_{a\pedex{lat}}| >  0.2\%$\footnotemark[3] &   3.07 &   3.29 &   0.61 \\
      \hline
      average $\delta_{B_0}$ (\%)     &   0.24 &   0.14 &  -0.27 \\
      rms average $\delta_{B_0}$ (\%) &   1.26 &   0.96 &   1.03 \\
      \% of materials with $|\delta_{B_0}| >  5.0\%$\footnotemark[1] &   4.29 &  55.26 &  55.21 \\
      \% of materials with $|\delta_{B_0}| >  5.0\%$\footnotemark[2] &   1.84 &   4.61 &   9.20 \\
      \% of materials with $|\delta_{B_0}| >  5.0\%$\footnotemark[3] &   0.61 &   0.00 &   0.00 \\
      \hline
      total number of materials &    163 &    152 &    163 \\
    \end{tabular}
    \footnotetext[1]{With an energy cutoff of $40 \text{Ry}$.}
    \footnotetext[2]{With an energy cutoff of $60 \text{Ry}$.}
    \footnotetext[3]{With an energy cutoff of $160 \text{Ry}$.}
  \end{ruledtabular}
\end{table}

For the zincblende compounds (cf. \mytabref{tab:result-zincblende}), we observe
results similar to for the rock-salt compounds.  We find that the USPPs
converge for most materials with an energy cutoff of $40\unit{Ry}$, whereas a
third of the materials with ONCV PP and half of the materials with PAW need an
energy cutoff of $60\unit{Ry}$ to converge.  Overall the accuracy of the ONCV
PP is slightly better than the alternatives, but all pseudoizations are on
average well below the target of $0.2\%$.  For the bulk moduli a larger energy
cutoff is necessary, but when converged the deviation from the AE results is
around $1\%$.  In \myfigref{fig:test}, we identify that only for BeO the
deviation between the ONCV calculation and the AE result is larger than
$0.2\%$.

\begin{figure}
  \includegraphics{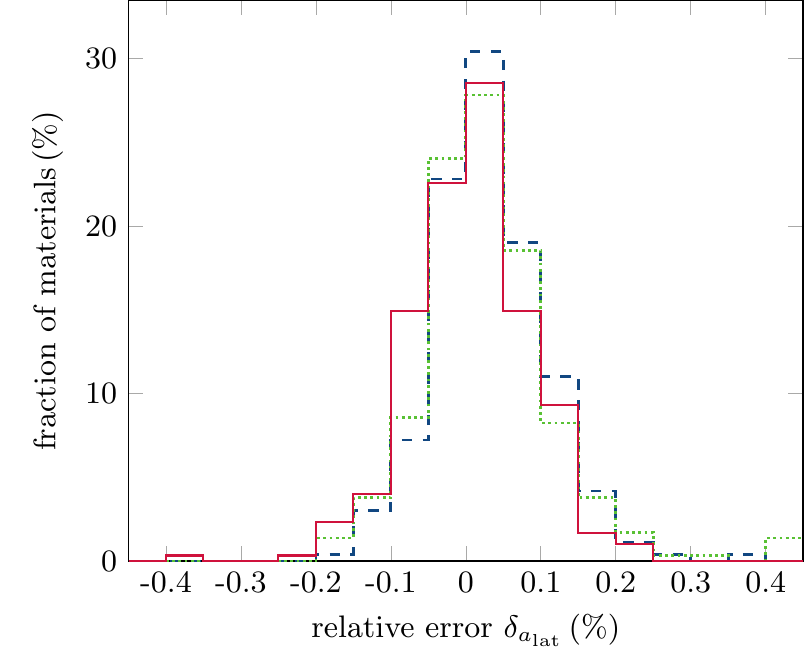}
  \centering

  \caption{(Color online) Histogram of the relative error of the lattice
constant compared to the all-electron result. We show the results for all
materials in the test set for the SG15 (red solid line), the GBRV (green dotted
line), and the PSLIB (blue dashed line) calculations.}

  \label{fig:histo-test}
\end{figure}

In \myfigref{fig:histo-test}, the histogram of the relative error of the
lattice constant for the test set confirms the conclusions we drew from
\mytabref{tab:result-sc} to \ref{tab:result-zincblende}: The deviation from the
all electron results is very small for all PP.  The USPP shows a slightly larger
deviation than the PAW and the ONCV PP.  The histogram reveals that this is
partly due to some outliers, for which the lattice constant is overestimated by
more than $0.4\%$.  Overall, we notice that the accuracy of the ONCV PP for the
test set of materials is not significantly worse than for the training set.
Hence, we are confident that these PP are transferable to other materials as
well.

\subsection{Dimers and ternary compounds}
\begin{table}
  \caption{Bond length of diatomic molecules and lattice constant of perovskites and half-Heusler compounds
  investigated with different methods. For the half-Heusler compounds, the first element is in Wyckoff position $c$. All values are given in \AA.}
  \label{tab:more-material}
  \begin{ruledtabular}
  	\begin{tabular}{c c c c c}
  	material & ref.\footnote{We evaluate the lattice constant perovskites and half Heusler with FLAPW and take the bond length of the dimers from the CCCB DataBase.\cite{cccbdb}} & GBRV & PSLIB & ONCV \\
  	\hline
  	H\pedex{2}  & 0.750 & 0.757 & 0.750 & 0.749 \\
  	N\pedex{2}  & 1.102 & 1.108 & 1.110 & 1.101 \\
  	O\pedex{2}  & 1.218 & 1.224 & 1.230 & 1.221 \\
  	F\pedex{2}  & 1.412 & 1.424 & 1.419 & 1.417 \\
  	Cl\pedex{2} & 2.012 & 2.004 & 2.006 & 2.015 \\
  	Br\pedex{2} & 2.311 & 2.311 &       & 2.314 \\
  	\hline
  	AsNCa\pedex{3} & 4.764 & 4.765 & 4.764 & 4.764 \\
  	BaTiO\pedex{3} & 4.018 & 4.028 & 4.029 & 4.020 \\
  	KMgCl\pedex{3} & 5.024 & 5.023 & 5.025 & 5.023 \\
  	LaAlO\pedex{3} & 3.814 & 3.817 & 3.815 & 3.809 \\
  	PNCa\pedex{3}  & 4.720 & 4.720 & 4.720 & 4.719 \\
  	SrTiO\pedex{3} & 3.937 & 3.939 & 3.942 & 3.938 \\
  	\hline
    BScBe  & 5.318 & 5.319 & 5.316 & 5.317 \\
    GeAlCu & 5.910 & 5.914 & 5.913 & 5.920 \\
    NiScSb & 6.118 & 6.120 & 6.123 & 6.123 \\
    NMgLi  & 5.004 & 5.006 & 5.006 & 5.010 \\
    PdZrSn & 6.392 & 6.392 & 6.394 & 6.394 \\
    PZnNa  & 6.141 & 6.149 & 6.148 & 6.148 \\
  	\end{tabular}
  \end{ruledtabular}
\end{table}

Our training and test set are limited to mono- and diatomic crystals, hence one
may wonder if the constructed ONCV PPs work outside this scope.  To test this
we investigated diatomic molecules and ternary compounds.  For the compounds,
we use the same computational setup as for the materials in the training and in
the test set.  For the molecules, we optimize the bond length inside a box with
dimensions $15\unit{\AA}\times 15\unit{\AA}\times 30\unit{\AA}$ with the long side parallel
to the axis of the molecule.

In \mytabref{tab:more-material}, we show the bond lengths and the lattice
constants of the investigated materials.  Depending on the pseudoization, some
diatomic molecules show large deviations from the reference data from the
CCCB~DataBase.\cite{cccbdb} Overall, the ONCV PPs exhibit the smallest
deviations.  The relative error is larger than $0.2\%$ only for the O\pedex{2}
(0.25\%) and the F\pedex{2} (0.35\%) dimer.  For the USPP, all diatomic
molecules are outside of the desired relative accuracy of $0.2\%$, except for
the Br\pedex{2} dimer.  In PAW, the only molecule with the desired accuracy is
the H\pedex{2} dimer.  The other molecules show deviations of similar magnitude
to the USPP and the Br\pedex{2} dimer did not converge.

Perovskites are accurately described by all pseudoizations; we frequently find
a relative agreement of better than $0.1\%$ in the lattice constant with the
FLAPW result.  The worst case for the ONCV PP is LaAlO\pedex{3}, which deviates
by $-0.13\%$.  The USPP and the PAW both overestimate the lattice constant of
BaTiO\pedex{3} by $0.25\%$ and $0.27\%$, respectively.  The PAW potentials also
feature a larger deviation than the other two pseudoizations for
SrTiO\pedex{3}.

Finally, we consider the half-Heusler compounds.  All materials are within the
desired accuracy with all pseudoizations.  The ONCV PP show slightly larger
deviations than USPP and PAW for GeAlCu and NMgLi.  For NiScSb, the ONCV PP and
PAW overestimate the lattice constant more than the USPP.  The lattice constant
of BScSb and PdZrSn are essentially the same with FLAPW and in any
pseudoization used.  In PZNa, all PP produce very similar results and a
slightly larger lattice constant than the FLAPW result.

\section{Conclusion} \label{sec:concl}

\begin{table}
  \caption{Summary of the results depicted in \mytabref{tab:result-bcc} to \ref{tab:result-zincblende} with 
    same notation as \mytabref{tab:result-bcc}.}
  \label{tab:result-summary}
  \begin{ruledtabular}
    \begin{tabular}{c c c c}
      & GBRV & PSLIB & SG15 \\
      \hline
      average $\delta_{a\pedex{lat}}$ (\%)     &   0.03 &   0.03 &   0.01 \\
      rms average $\delta_{a\pedex{lat}}$ (\%) &   0.12 &   0.09 &   0.08 \\
      \% of materials with $|\delta_{a\pedex{lat}}| >  0.2\%$\footnotemark[1] &   7.04 &  38.51 &  30.07 \\
      \% of materials with $|\delta_{a\pedex{lat}}| >  0.2\%$\footnotemark[2] &   6.70 &  10.22 &   4.65 \\
      \% of materials with $|\delta_{a\pedex{lat}}| >  0.2\%$\footnotemark[3] &   6.19 &   3.73 &   1.99 \\
      \hline
      average $\delta_{B_0}$ (\%)     &   0.21 &   0.18 &  -0.14 \\
      rms average $\delta_{B_0}$ (\%) &   2.61 &   2.85 &   2.40 \\
      \% of materials with $|\delta_{B_0}| >  5.0\%$\footnotemark[1] &  13.75 &  60.51 &  54.49 \\
      \% of materials with $|\delta_{B_0}| >  5.0\%$\footnotemark[2] &   7.73 &  13.36 &  13.62 \\
      \% of materials with $|\delta_{B_0}| >  5.0\%$\footnotemark[3] &   4.47 &   3.34 &   3.82 \\
      \hline
      total number of materials &    582 &    509 &    602 \\
    \end{tabular}
    \footnotetext[1]{With an energy cutoff of $40 \text{Ry}$.}
    \footnotetext[2]{With an energy cutoff of $60 \text{Ry}$.}
    \footnotetext[3]{With an energy cutoff of $160 \text{Ry}$.}
  \end{ruledtabular}
\end{table}

We have presented an algorithm to optimize the input parameters of a
pseudopotential (PP) construction.  We demonstrated it by developing the SG15
dataset\cite{webpage} of ONCV pseudopotentials, which exhibits a similar accuracy as the
ultrasoft PP database GBRV\cite{gbrv14} and the PAW library PSLIB.\cite{cors14}
The idea of the algorithm is to map the PP onto a single numeric value so that
standard optimization techniques can be employed.  For this, we developed a
quality function that considers the accuracy of the lattice constant of a PP
calculation and compares it with a high accuracy FLAPW one.  In addition, the
quality function takes into account the energy cutoff necessary to converge the
calculation.  Hence, the optimzation of the PP with respect to the quality
function yields accurate and efficient potentials.  In order to ensure that the
constructed PPs are of a high accuracy, we systematically chose a set of
approximately 600 materials and evaluate their properties with FLAPW.  We split
this set in two parts, a training set used for the optimization of the PP and a
test set to analyze the performance of the PP.  When a PP does not produce our
desired accuracy after optimizing on the training set, we can improve the
quality of the PP by extending the training set by more materials.

In \mytabref{tab:result-summary}, we collect the results of all materials in
test and training set.  Compared to the PP from the GBRV database\cite{gbrv14}
and PSLIB,\cite{cors14} the PP in the SG15 set have the lowest root-mean-square
deviation from the FLAPW results for the lattice constant.  With an energy
cutoff of $60\unit{Ry}$, the ONCV PP feature the least number of materials with
an inaccurate lattice constant (deviation larger than 0.2\% from FLAPW
results).  The advantage of the ultrasoft PP is that they offer a similar
accuracy with an energy cutoff of $40\unit{Ry}$.  For the bulk moduli larger
energy cutoffs are necessary for all pseudoization methods.  The ONCV PP have
the smallest root-mean-square deviation for the tested materials.  The fraction
of materials that can be accurately described with the ONCV PP at a certain
energy cutoff is very similar to the performance of the PAW.  The ultrasoft PP
exhibit a similar accuracy at a moderately lower energy cutoff.  For materials
that go beyond the training and test set, we find that the ONCV PP provides the
best description of diatomic molecules.  All pseudopotentials are very accurate
for perovskite and half-Heusler compounds.

We encourage the community to use the algorithm presented in this work to
optimize pseudopotentials for different functionals and with different
construction methods.  With only a modest increase in the energy cutoff, the
proposed SG15 library of norm-conserving pseudopotentials provides a
competitive alternative to the libraries using USPP and PAW.  As these
pseudopotentials are less complex than the alternatives, this results in a
great simplification in the development and implementation of new algorithms.

\begin{acknowledgments}
This work was supported by the US Department of Energy through
grant DOE-BES DE-SC0008938.
An award of computer time was provided by the DOE Innovative and
Novel Computational Impact on Theory and Experiment
(INCITE) program. This research used resources of the
Argonne Leadership Computing Facility at Argonne National
Laboratory, which is supported by the Office of Science of the
U.S. Department of Energy under contract DE-AC02-06CH11357.
\end{acknowledgments}

\bibliographystyle{prsty}
\bibliography{database}

\begin{thebibliography}{10}

\bibitem{Martin2004}
See e.g. R. M. Martin, Electronic Structure. Basic Theory and Practical Methods, Cambridge University Press, 2004.

\bibitem{hk64}
P. Hohenberg and W. Kohn, Phys. Rev. {\bf 136},  B864  (1964).

\bibitem{ks65}
W. Kohn and L.~J. Sham, Phys. Rev. {\bf 140},  A1133  (1965).

\bibitem{hsc79}
D.~R. Hamann, M. Schl\"uter, and C. Chiang, Phys. Rev. Lett. {\bf 43},  1494
  (1979).

\bibitem{bhs82}
G.~B. Bachelet, D.~R. Hamann, and M. Schl\"uter, Phys. Rev. B {\bf 26},  4199
  (1982).

\bibitem{vand90}
D. Vanderbilt, Phys. Rev. B {\bf 41},  7892  (1990).

\bibitem{paw}
P.~E. Bl\"ochl, Phys. Rev. B {\bf 50},  17953  (1994).

\bibitem{hama13}
D.~R. Hamann, Phys. Rev. B {\bf 88},  085117  (2013).

\bibitem{bec2_93}
A.~D. Becke, J. Chem. Phys. {\bf 98},  1372  (1993); {\it ibid}. {\bf 98}, 5648
  (1993).

\bibitem{ag98}
F. Aryasetiawan and O. Gunnarsson, Rep. Prog. Phys. {\bf 61},  237  (1998).

\bibitem{bgt87}
S. Baroni, P. Giannozzi, and A. Testa, Phys. Rev. Lett. {\bf 58},  1861
  (1987).

\bibitem{tm91}
N. Troullier and J.~L. Martins, Phys. Rev. B {\bf 43},  1993  (1991).

\bibitem{hgh98}
C. Hartwigsen, S. Goedecker, and J. Hutter, Phys. Rev. B {\bf 58},  3641
  (1998).

\bibitem{gbrv14}
K.~F. Garrity, J.~W. Bennett, K.~M. Rabe, and D. Vanderbilt, Comp. Mater. Sci.
  {\bf 81},  446   (2014).

\bibitem{cors14}
A. {Dal Corso}, Comp. Mater. Sci. {\bf 95},  337   (2014).

\bibitem{wkwf81}
E. Wimmer, H. Krakauer, M. Weinert, and A.~J. Freeman, Phys. Rev. B {\bf 24},
  864  (1981).

\bibitem{wwf82}
M. Weinert, E. Wimmer, and A.~J. Freeman, Phys. Rev. B {\bf 26},  4571  (1982).

\bibitem{jf84}
H.~J.~F. Jansen and A.~J. Freeman, Phys. Rev. B {\bf 30},  561  (1984).

\bibitem{lsoc14}
K. Lejaeghere, V. Van~Speybroeck, G. Van~Oost, and S. Cottenier, Crit. Rev.
  Solid State Mater. Sci. {\bf 39},  1  (2014).

\bibitem{mur44}
F. Murnaghan, Proc. Nat. Acad. Sci. USA {\bf 30},  244  (1944).

\bibitem{kucu14}
E. Kucukbenli, M. Monni, B. Adetunji, X. Ge, G. Adebayo, N. Marzari, S.
  de~Gironcoli, and A.~D. Corso, arXiv:1404.3015  .

\bibitem{webpage}
http://www.quantum-simulation.org

\bibitem{pbe96}
J.~P. Perdew, K. Burke, and M. Ernzerhof, Phys. Rev. Lett. {\bf 77},  3865
  (1996).

\bibitem{fleur}
http://www.flapw.de.

\bibitem{QE2009}
P. Giannozzi, S. Baroni, N. Bonini, M. Calandra, R. Car, C. Cavazzoni, D.
  Ceresoli, G.~L. Chiarotti, M. Cococcioni, I. Dabo, A. {Dal Corso}, S.
  de~Gironcoli, S. Fabris, G. Fratesi, R. Gebauer, U. Gerstmann, C. Gougoussis,
  A. Kokalj, M. Lazzeri, L. Martin-Samos, N. Marzari, F. Mauri, R. Mazzarello,
  S. Paolini, A. Pasquarello, L. Paulatto, C. Sbraccia, S. Scandolo, G.
  Sclauzero, A.~P. Seitsonen, A. Smogunov, P. Umari, and R.~M. Wentzcovitch, J.
  Phys.: Condens. Matter {\bf 21},  395502 (19pp)  (2009).

\bibitem{nm65}
J.~A. Nelder and R. Mead, The Computer Journal {\bf 7},  308  (1965).


\bibitem{cccbdb}
{\em NIST Computational Chemistry Comparison and Benchmark Database, NIST
  Standard Reference Database Number 101, Release 16a}, edited by R.~D.
  {Johnson III} (August 2013).

\end{thebibliography}

\end{document}